\documentclass[twocolumn,floatfix,prl,superscriptaddress,nofootinbib]{revtex4-2}
\usepackage{enumitem}
\usepackage{pgfplots}
\pgfplotsset{compat=1.17}
\usepackage{graphicx}
\usepackage{amsmath}
\usepackage{amsfonts}
\usepackage{amsthm}
\usepackage{braket}
\usepackage{url}
\usepackage{numprint}
\usepackage{hyperref}
\usepackage[capitalise]{cleveref}
\usepackage{color}
\usepackage{algpseudocode}
\newenvironment{algorithm}[1][]{\begin{figure}[#1]}{\end{figure}}
\usepackage{tikz}
\usepackage{booktabs}
\usepackage[makeroom]{cancel}
\graphicspath{{Graphs/}}

\newtheorem{corollary}{Corollary}
\newtheorem{lemma}{Lemma}
\newtheorem{proposition}{Proposition}
\newtheorem{remark}{Remark}

\crefname{corollary}{corollary}{corollaries}
\Crefname{corollary}{Corollary}{Corollaries}

\begin{document}

\title{Quantum-Assisted Counterfactual Regret Minimization for Bounded-Payoff Imperfect-Information Games}

\author{G.~Maresch}
\affiliation{TU Wien, Institute of Discrete Mathematics and Geometry, Wiedner Hauptstra{\ss}e 8/104, Vienna, Austria}
\author{E.~Schulze}
\affiliation{Bull GmbH, von-der-Wettern-Straße 27, 51149 Cologne, Germany}
\author{J.~Yalla}
\affiliation{DataTeach.ai, Hyderabad, India}

\begin{abstract}

Imperfect-information games with a graded-reward structure present significant computational challenges for equilibrium computation, since the payoff expectations must be estimated across exponentially large strategy spaces. \textit{Counterfactual Regret Minimization} over the years has proven effective for approximating Nash equilibria, but Monte Carlo variants create a large overhead of sampling costs when the counterfactual values are estimated from sampled game-tree trajectories. We propose a novel Quantum-Assisted Counterfactual Regret Minimization, a framework that uses Quantum Amplitude Estimation to reduce the computational complexity of the regret updates in bounded-payoff imperfect-information games, while preserving the convergence of its classical counterpart. We give a theoretical analysis of the resulting exploitability bound, discuss NISQ-compatible amplitude-estimation variants, the structure of their resource requirements, and outline what is needed to validate the potential advantage empirically.

\end{abstract}

\maketitle

\section{Introduction}

Imperfect-information games, such as poker or security-resource-allocation problems, are a fundamental challenge in computational game theory because of the exponentially growing strategy spaces and multiple agents acting under hidden information. Many of our current economically and strategically relevant decisions are made under these conditions of uncertainty against adaptive opponents: bidding in auctions, negotiating contracts, allocating limited defensive resources against an attacker, or pricing. The quality of a decision is measured by how much a best-responding opponent could gain against it. Computing strategies is therefore a central problem of computational game theory and a building block for decision-making systems in auctions and market design, cybersecurity and physical-security planning, and automated negotiation. Counterfactual Regret Minimization (CFR) \cite{Zinkevich} is the workhorse for this task: it scales to games with very large instances and comes with convergence guarantees. The first demonstration that imperfect-information games of realistic size can be solved at a level beyond top human players was done by Brown \emph{et al.}\  in poker \cite{Brown}. The computational bottleneck of CFR is the repeated estimation of counterfactual values, which in large games is done by Monte Carlo sampling and scales as $\mathcal{O}(\varepsilon^{-2})$ in the target precision $\varepsilon$. Since this estimation is an expectation-value problem over a bounded payoff, it is just the type of task for which quantum amplitude estimation offers a quadratic improvement in precision-dependent query complexity \cite{Grinko2019,Suzuki2019}. This motivates embedding amplitude estimation into the regret-update loop of CFR as an iterative algorithm for approximating Nash equilibria in such games.  Vanilla CFR computes counterfactual values by traversing the game tree, whereas Monte Carlo CFR (MCCFR) variants reduce the computational cost per iteration by sampling portions of the game tree.

Unlike prior quantum game-theoretic work, which largely targets zero-sum games with binary or fixed payoffs \cite{Vasconcelos}, many practically relevant imperfect-information settings, such as graded bidding, multi-round negotiation, security games with continuous loss functions have graded rewards distributed over a continuum rather than a small discrete outcome set. Graded rewards do not intrinsically imply higher variance, but they naturally lead to counterfactual-value estimation as a general bounded-payoff expectation-estimation problem, to which QAE applies directly.

Our approach is structured as follows: We (i) formally describe how QAE is applied to counterfactual-value computation within CFR, (ii) derive an exploitability bound for the resulting average strategy that separates the classical $\mathcal{O}(T^{-1/2})$ convergence term from an additive $\mathcal{O}(\varepsilon)$ quantum-estimation-error term, (iii) discuss NISQ-compatible QAE variants (IQAE, ML-AE), (iv) analyze the effect of realistic gate, readout, and shot noise on the crossover point at which quantum estimation becomes advantageous. We close by identifying the concrete simulation and hardware experiments still needed before an empirical quantum advantage can be claimed.

\section{Related Work}
\label{sec:related}

\subsection{Classical equilibrium computation and CFR}

Counterfactual Regret Minimization \cite{Zinkevich} reduces the approximation of Nash-equilibria in an extensive-form game to the following: at every information set, each action's regret is accumulated over iterations and future play is chosen to be proportional to positive regret. The time-averaged strategy profile converges to an $\varepsilon$-Nash equilibrium at rate $\mathcal{O}(T^{-1/2})$. Sampling variants of CFR (chance-sampling, outcome-sampling, external-sampling, and more recent variance-reduced and deep variants) trade an increase in per-iteration variance for a reduction in per-iteration work \cite{Brown}. All of these methods remain fundamentally limited by the $\mathcal{O}(\varepsilon^{-2})$ scaling of classical Monte Carlo estimation.

\subsection{Quantum approaches to game-theoretic equilibria}

Vasconcelos \emph{et al.}\ \cite{Vasconcelos} study \emph{quantum zero-sum games}, in which two players submit unentangled quantum states to a referee, and introduce the Optimistic Matrix Multiplicative Weights Update (OMMWU) algorithm. OMMWU achieves $\mathcal{O}(d \varepsilon^{-1})$ iterations to an $\varepsilon$-Nash equilibrium, what is a quadratic improvement in $\varepsilon$ over the $\mathcal{O}(d \varepsilon^{-2})$ rate of the original Matrix Multiplicative Weights Update algorithm of Jain and Watrous \cite{JainWatrous2009}. The authors further assume that this rate is optimal within the class of optimistic matrix-multiplicative-weight (OMMP-type) methods. More specifically, they conjecture that no algorithm in this class can improve on $\mathcal{O}(\varepsilon^{-1})$ iterations for all instances. Subsequent work has shown that the $1/\varepsilon$ barrier can in fact be overcome under additional structural assumptions in the game, so the conjecture should be read as characterizing the OMMP family (rather than as a more universal lower bound).

This line of work targets the iteration complexity in games whose payoffs are themselves quantum-mechanical (measurement outcomes on shared quantum states). Q-CFR is complementary: the game itself is a classical extensive-form game with hidden information and classical payoffs. The quantum resource is used only as a subroutine for estimating expectations that would otherwise require Monte Carlo sampling. The two notions of ``quantum game'' (games whose payoff structure is quantum), versus classical games (solvers use quantum subroutines) should not be conflated.

\subsection{Quantum subroutines in adjacent equilibrium formulations}

Closest to our setting is the use of a quantum subroutine, such as Quantum Counting (that includes QPE) for the quantum implementation of the imperfect-information game of Skat \cite{Skat}. There, the game state is encoded in quantum
registers, gates implement the game progress and rules, and a score operator
projects onto winning outcomes.

This shows that amplitude-estimation-type primitives can evaluate the game tree of an imperfect-information game. Nevertheless, Q-CFR differs from that approach by accumulating regrets to converge to Nash equilibria, instead of evaluating the decisions of a single position.

We also note prior work that uses Grover search as a subroutine \emph{within} a fixed probabilistic risk-propagation model to identify sensitive parameters \cite{Jos}. We regard it as evidence that quantum subroutines are, in general, reusable primitives.
In that particular work, the quantum subroutines serve as an analysis tool within a fixed risk model. The work itself is not concerned with strategic decision-making or equilibrium computation, in contrast to Q-CFR.

\subsection{Quantum amplitude estimation}

Canonical QAE \cite{Brassard} combines Grover iterations with Quantum Phase Estimation (QPE) to estimate an amplitude $a$ to additive error $\varepsilon$ using only $\mathcal{O}(\varepsilon^{-1})$ oracle calls. Canonical QAE requires $\mathcal{O}(\log(1/\varepsilon))$ auxiliary qubits for QPE and correspondingly deep circuits, which makes it impractical on near-term devices. This motivated QPE-free variants, in particular Iterative QAE (IQAE) \cite{Grinko2019} and Maximum-Likelihood Amplitude Estimation (ML-AE) \cite{Suzuki2019}, both of which retain the $\mathcal{O}(\varepsilon^{-1})$ query scaling (up to logarithmic factors) while removing the need for QPE auxiliary qubits. We build on these NISQ-compatible variants in Sec.~\ref{sec:nisq}.

\section{The Q-CFR Framework}
\label{sec:qcfr-algorithm}

\begin{algorithm}[htbp]
\hrule
\vspace{2pt}

\caption{Quantum-Assisted Counterfactual Regret Minimization (Q-CFR)}

\vspace{-2pt}
\hrule
\vspace{4pt}

\label{alg:qcfr}
\begin{algorithmic}[1]
\Procedure{Q-CFR}{$G$, $\varepsilon$, $\alpha$, $T$}
    \State Initialize $R^0(I,a) \gets 0$ for all $(I,a)$

    \For{$t = 1$ to $T$}
        \State \textbf{Strategy update (regret matching):}
        \ForAll{$(I, a)$}
            \If{$\sum_{b \in A(I)} \max(R^{t-1}(I,b),0) > 0$}
                \State $\displaystyle
                \sigma^t(I,a) \gets
                \frac{\max(R^{t-1}(I,a),0)}
                {\sum_{b \in A(I)}\max(R^{t-1}(I,b),0)}$
            \Else
                \State $\displaystyle
                \sigma^t(I,a) \gets \frac{1}{|A(I)|}$
            \EndIf
        \EndFor
        \State 
        \State \textbf{Quantum-oracle call:}
        \ForAll{$(I,a)$}
            \State Estimate $\tilde v^{\sigma^t}(I,a)$ such that
            \Statex \hspace{\algorithmicindent}\hspace{\algorithmicindent}\hspace{\algorithmicindent}
            $\Pr\left(
            |\tilde v^{\sigma^t}(I,a)-v^{\sigma^t}(I,a)| \le \varepsilon
            \right) \ge 1-\alpha$
        \EndFor

        \ForAll{$I$}
            \State $\displaystyle
            \tilde v^{\sigma^t}(I)
            \gets \sum_{a\in A(I)}
            \sigma^t(I,a)\tilde v^{\sigma^t}(I,a)$
        \EndFor
        \State
        \State \textbf{Regret update:}
        \ForAll{$(I,a)$}
            \State $r^t(I,a) \gets
            \tilde v^{\sigma^t}(I,a)-\tilde v^{\sigma^t}(I)$
            \State $R^t(I,a) \gets R^{t-1}(I,a)+r^t(I,a)$
        \EndFor

    \State
    \State \textbf{Compute player reach probability}  $\displaystyle \pi^{\sigma^t}(I)$
    \State
    \EndFor

    \State \Return $\displaystyle
    \bar{\sigma}_T(I,a)=
    \frac{\sum_{t=1}^{T}\pi^{\sigma^t}(I)\sigma^t(I,a)}
         {\sum_{t=1}^{T}\pi^{\sigma^t}(I)}$
\EndProcedure
\end{algorithmic}
\vspace{2pt}
\hrule
\end{algorithm}

We need to estimate the total error that accumulates if we replace the CFR-values $v^t(a, I)$ by a quantum-assisted estimate $\hat{v}^t(a, I)$. From this an estimator for the instantaneous regret $r^t(I,a)$ is obtained. The value-estimation error $e^t(a,I)=v^t(a,I)-\hat{v}^t(a, I)$ is treated as a probabilistic random variable in \cite{gibson2012generalized}, Lemma 1.

Let us first collect some facts about $e^t(a,I)$:

\begin{lemma}
For fixed information set $I$ and action $a$, let both $v^t(a,I), \hat{v}^t(a, I) \in [-1,1]$. If for $e^t(a,I)=v^t(a,I)-\hat{v}^t(a, I)$
\begin{align}
    \mathbb{P}[|e^t(a,I)|\le \varepsilon] \ge 1 - \alpha
\end{align}
then the variance $\operatorname{Var}(e^t(a,I))$ is bounded $\le (1-\alpha)\varepsilon^2 + 4\alpha$ and $\mathbb{E} (|e^t(a,I)|)$ is bounded by $ \le (1-\alpha)\varepsilon + 2\alpha$. In particular if $\alpha$ is chosen as $\mathcal{O}(\varepsilon^2)$, then the variance is also $\mathcal{O}(\varepsilon^2)$.\end{lemma}
\begin{proof} It suffices to note that $|e^t(a,I)|\le 2$. Then Chebychev's inequality immediately implies the bound on bias. The bound on the variance follows by the trivial estimate $\operatorname{Var}(e^t(a,I))\le \mathbb{E} (|e^t(a,I)^2|)$.
\end{proof}

Ultimately we are interested in $d^t(a,I)=r^t(a,I)-\hat{r}^t(a, I)$ where 
\begin{align}
r^t(a,I)&=v^t(a,I)-\sum_b\sigma_t(I,b) v_t(I,b),
\end{align}
 for the actions available at $I$ and the current strategy at iteration $t$.

We will need an estimate on the covariance $\operatorname{Cov}(d^s,d^t)$ for $t\neq s$:

\begin{lemma}
If for times $t,s$ and all then available information sets $I$ and actions $a$ uniformly
\begin{align}
    \mathbb{P}[|e^t(a,I)|\le \varepsilon] \ge 1 - \alpha.
\end{align} Then for $d^t$ defined as above, $\operatorname{Var}(d^t)$ and $|\operatorname{Cov}(d^s, d^t)|$ both are bounded by $4(1-\alpha)\varepsilon^2 + 16\alpha$.
\end{lemma}

\begin{proof} 
Going through the definitions we can write 
\begin{align}
    d^t(a,I) &= e_t(a,I) - \sum_b\sigma_t(I,b) e^t(I,b).
\end{align}
For convenience let us abbreviate $q_{\varepsilon, \alpha}=(1-\alpha)\varepsilon^2 + 4\alpha$. Then, by the triangle inequality for square-integrable random variables
\begin{align}
    \sqrt{\mathbb{E}(d^t(a,I)^2)} &\le \sqrt{\mathbb{E}(e^t(a,I)^2)} + \sum_b|\sigma_t(I,b)|  \sqrt{\mathbb{E}(e^t(b,I)^2)}.
\end{align}
The strategy weights $\sigma_t(I,\cdot)$ are non-negative and sum to $1$, therefore invoking the uniform variance bound on the $e^t(a,I)$ terms, one gets
\begin{align}
    \sqrt{\mathbb{E}(d^t(a,I)^2)} &\le \sqrt{q_{\varepsilon, \alpha}} + \sqrt{q_{\varepsilon, \alpha}},
\end{align}
which in turn implies $\operatorname{Var}(d^t(a,I)) \le 4 q_{\varepsilon, \alpha} $.
Now it is enough to apply the Cauchy-Schwarz inequality $|\operatorname{Cov}(d^s, d^t)| \le \sqrt{\operatorname{Var}(d^s)\cdot \operatorname{Var}(d^t)}$. 
\end{proof}

\begin{lemma}[Gibson et al.~\cite{gibson2012generalized}]
Let $p \in (0,1]$ and suppose there exists a bound
$\widehat{\Delta}$ on the difference between any two
estimated counterfactual values for player $i$, such that
\begin{align}    
    \hat{v}_i(a, I) - \hat{v}_i(b, I) \leq \widehat{\Delta}_i .
\end{align}
If the strategies are chosen by regret matching using the
estimated counterfactual regrets, then, with probability at
least $1-p$, the true average regret satisfies
\begin{align} 
    \frac{R_i^T}{T} \leq
    |\mathcal{I}_i|\sqrt{|A_i|}\left(
 \frac{\widehat{\Delta}_i}{\sqrt{T}} + \sqrt{\frac{\operatorname{Var}}{pT}+\frac{\operatorname{Cov}}{p}+\frac{E^2}{p}}\right),
\end{align}
where
\begin{align}
    \operatorname{Var} =
    \max_{\substack{
        t \in \{1,\ldots,T\}\\
        I \in \mathcal{I}_i\\
        a \in A(I)
    }}
    \operatorname{Var}
    \left[
        r_i^t(I,a)
        -
        \widehat{r}_i^t(I,a)
    \right],    
\end{align}

and $\operatorname{Cov}$ and $E$ are defined analogously as maxima over all iterations, information sets and actions.
\end{lemma}
If we enter our bounds for variances and covariances of the QAE assisted approximation we immediately get
\begin{corollary}
    If uniformly for all times $t\le T$ and all information sets $I$ and actions $a$
\begin{align}
    \mathbb{P}[|e^t(a,I)|\le \varepsilon] \ge 1 - \alpha.
\end{align} Then with probability at least $1-p$
\begin{align}
    \frac{R_i^T}{T} \leq 2 |\mathcal{I}_i|\sqrt{|A_i|}\left(
 \frac{1}{\sqrt T}+\sqrt{\frac{q_{\varepsilon, \alpha}}{p} \left(2+\frac{1}{T}\right)}
\right).
\end{align}
where $q_{\varepsilon, \alpha}=(1-\alpha)\varepsilon^2 + 4\alpha$. In particular, for $\alpha = \mathcal{O}(\varepsilon^2)$ the true average regret is $\mathcal{O}(T^{-1/2} + \varepsilon)$ with confidence level $\ge 1-p$
\end{corollary}
For a two-player zero-sum game the standard exploitability bound in terms of the true average regret $\operatorname{Exploitability}(\bar{\sigma}^T)
\le \tfrac{1}{2} \sum_{i=1}^2
\frac{R_i^T}{T}$ directly implies the following 
\begin{corollary}
\label{thm:qcfr_bound}
For a finite two-player zero-sum extensive-form game
with perfect recall under the assumptions above, with probability at least $1-2p$, the $\operatorname{Exploitability}(\bar{\sigma}^T)$ is bounded by
\begin{align}
    \sum_{i=1}^{2}
    |\mathcal{I}_i|\sqrt{|A_i|}
    \left(
        \frac{1}{\sqrt{T}}
        +
        \sqrt{
            \frac{q_{\varepsilon,\alpha}}{p}
            \left(2+\frac{1}{T}\right)
        }
    \right)
\end{align}
where $q_{\varepsilon, \alpha}=(1-\alpha)\varepsilon^2 + 4\alpha$. In particular, for $\alpha=\mathcal{O}(\varepsilon^2)$ and fixed $p$,
\begin{align}
    \operatorname{Exploitability}(\bar{\sigma}^T)
    =
    \mathcal{O}\!\left(\frac{1}{\sqrt{T}}+\varepsilon\right),
\end{align}
up to game-dependent constants.
\end{corollary}

In future work it will prove interesting to obtain empirical variance and covariance bounds instead of the worst case estimates used above.
An important question is whether the QAE estimator is in practice indeed unbiased. If so, one could use even sharper bounds, cf. \cite[Theorem 2]{gibson2012generalized}.

\begin{corollary}
\label{cor:speedup}
Let $N_{\mathcal G}$ denote the number of counterfactual action values estimated in one Q-CFR iteration. Suppose estimating the counterfactual value $v^t(a, I)$ up to error $\varepsilon$ with probability at least $1-\alpha$ requires
$\mathcal{O}(\varepsilon^{-2}\log(1/\alpha))$ Monte Carlo samples, whereas QAE
requires
$\mathcal{O}(\varepsilon^{-1})$ oracle calls up to logarithmic terms.

For a fixed game and fixed confidence parameter $p$, the exploitability bound of Corollary~\ref{thm:qcfr_bound} is
$\mathcal{O}(T^{-1/2}+\sqrt{q_{\varepsilon,\alpha}})$, with
$q_{\varepsilon,\alpha} = (1-\alpha)\varepsilon^2+4\alpha$. Hence, for given target exploitability $\delta$, the other parameters
\begin{align}
\varepsilon=\Theta(\delta),
\qquad
\alpha=\Theta(\delta^2),
\qquad
T=\Theta(\delta^{-2}),
\end{align} can be chosen such that  $q_{\varepsilon,\alpha}=\mathcal{O}(\delta^2)$.

The total estimation costs are therefore up to logarithmic contributions
$\mathcal{O}\left(\delta^{-3}N_{\mathcal G}\right)$
QAE oracle calls and $\mathcal{O}\!\left(\delta^{-4}N_{\mathcal G} \right)$
Monte Carlo samples.
\end{corollary}
Thus QAE gives a quadratic improvement in the
precision-dependent query complexity of counterfactual-value
estimation. This is a query-complexity statement, not a claim of an
end-to-end quadratic runtime speedup over classical CFR.

\begin{remark}
From \Cref{thm:qcfr_bound} one gets that the asymptotic exploitability satisfies $\mathrm{Exploitability}(\bar\sigma^T) = \mathcal{O}\left(T^{-1/2} + \varepsilon\right)$: Q-CFR retains the classical $\mathcal{O}(T^{-1/2})$ convergence rate of CFR, with quantum estimation contributing only an additive, tunable $\mathcal{O}(\varepsilon)$ term that vanishes as more precise (and correspondingly more expensive) QAE calls are used.
\end{remark}

\section{NISQ-Friendly QAE Variants}
\label{sec:nisq}

Canonical QAE requires $\mathcal{O}(\log(1/\varepsilon))$ auxiliary qubits for quantum phase estimation and correspondingly deep circuits, which makes it impractical on near-term devices. Two QPE-free variants are of particular interest for Q-CFR.

\textbf{Iterative Quantum Amplitude Estimation (IQAE)} Grinko \emph{et al.}\ \cite{Grinko2019} replace QPE with an adaptive sequence of Grover-operator powers $k_i$, chosen from previous measurement outcomes, and a classical confidence-interval update. IQAE needs no auxiliary qubits beyond those required for problem encoding, and its circuit depth scales as $\mathcal{O}(\varepsilon^{-1})$, matching canonical QAE's oracle-call scaling while using substantially fewer qubits. Its adaptive structure mirrors the iterative nature of CFR's own regret updates, which makes it a natural default choice for Q-CFR.

\textbf{Maximum-Likelihood Amplitude Estimation (ML-AE)} \cite{Suzuki2019} instead uses a fixed schedule of Grover powers (typically $\{2^j\}_{j=0}^{m}$) and recovers the amplitude via classical maximum-likelihood post-processing of the resulting measurement statistics. This trades a predetermined, hardware-schedulable set of circuit depths for additional classical post-processing, which can simplify resource planning when many counterfactual values must be estimated in parallel across a batch of information sets.

Both variants achieve $\mathcal{O}(\varepsilon^{-1})$ query scaling up to logarithmic factors, so the choice between them is primarily one of qubit budget (favoring IQAE) versus scheduling predictability (favoring ML-AE), rather than one of asymptotic sample complexity.

\subsection{Shot budgets and confidence}

In practice each Grover-power step of IQAE (or each fixed depth of ML-AE) is itself repeated for $N_{\mathrm{shots}}$ measurement shots to build the empirical estimate of the corresponding success probability, and the target additive error $\varepsilon$ and failure probability $\alpha$ jointly determine the total number of shots via the standard IQAE/ML-AE confidence-interval bounds. Because Q-CFR queries a counterfactual value $(\tilde v^{\sigma^t}(I,a)$ at every information set and every iteration, the total shot budget for a full run is the product of (i) the number of iterations $T$, (ii) the number of information-set/action pairs queried per iteration, and (iii) the per-call shot count implied by $\varepsilon$ and $\alpha$. This product, rather than $T$ alone, is the quantity that should be compared against the corresponding classical Monte Carlo rollout budget when assessing whether a given game instance is within reach of a quantum-advantage crossover (Sec.~\ref{sec:noise}).

\section{Noise Analysis and the Quantum-Advantage Crossover}
\label{sec:noise}

\subsection{Error model}

Under realistic noise, the effective QAE estimation error is a sum of several contributions,
\begin{align}
\varepsilon_{\text{total}} = \varepsilon_{\text{shot}} + \varepsilon_{\text{gate}} + \varepsilon_{\text{readout}} + \varepsilon_{\text{crosstalk}},
\end{align}
with $\varepsilon_{\text{shot}} = \mathcal{O}(N_{\text{shots}}^{-1/2})$ from finite sampling, $\varepsilon_{\text{gate}} = \mathcal{O}(p_{\text{gate}} D)$ from gate infidelity accumulated over circuit depth $D$, $\varepsilon_{\text{readout}} = \mathcal{O}(p_{\text{readout}})$ from measurement error, and $\varepsilon_{\text{crosstalk}} = \mathcal{O}(p_{\text{crosstalk}} N_{\text{qubits}})$ from residual inter-qubit coupling. For representative current superconducting-qubit error rates ($p_{\text{gate}} \sim 10^{-3}$ per two-qubit gate, $p_{\text{readout}} \sim 10^{-2}$), the gate-error term dominates once circuit depth exceeds a few tens of layers. This bounds the smallest $\varepsilon$ that can be reached, before noise (rather than shot count) becomes the limiting factor.

\subsection{Crossover condition}

Quantum amplitude estimation is advantageous over classical Monte Carlo whenever the corresponding query counts satisfy
\begin{align}
\frac{\mathcal{O}(\varepsilon_{\text{classical}}^{-2})}{\mathcal{O}(\varepsilon_{\text{quantum}}^{-1})} \cdot \frac{1}{1 + \varepsilon_{\text{noise}}/\varepsilon_{\text{ideal}}} > 1,
\end{align}
i.e., whenever
\begin{align}
\varepsilon<\varepsilon_{\text{crossover}} := \frac{C}{1 + \varepsilon_{\text{noise}}/\varepsilon_{\text{ideal}}}
\end{align} for some implementation dependent constant $C$.
Because $\varepsilon_{\text{gate}}$ grows with circuit depth while the target $\varepsilon_{\text{quantum}}$ shrinks as $1/D$, there is a depth (equivalently, a target precision) beyond which additional Grover iterations no longer improve the \emph{effective} estimation error, and the crossover condition above fails. Locating this depth for a specific device's calibration data is a prerequisite for any claim of practical quantum advantage and is identified in Sec.~\ref{sec:conclusion} as required future work, rather than assumed here.

\begin{proposition}

Q-CFR retains the classical CFR convergence behavior, in the sense that
the estimation-error contribution does not dominate the
$\mathcal{O}(T^{-1/2})$ term, whenever the total noise-induced estimation error
satisfies
\begin{align}
q_{\varepsilon_{\text{total}},\alpha}
\leq
\frac{p}{2T+1},
\end{align}
where
\begin{align}
q_{\varepsilon_{\text{total}},\alpha}
=
(1-\alpha)\varepsilon_{\text{total}}^2+4\alpha.
\end{align}
For fixed $p$ and sufficiently small $\alpha$, this corresponds to
$\varepsilon_{\text{total}}=\mathcal{O}(T^{-1/2})$, i.e., the noise floor must
stay below the classical CFR convergence rate at the chosen iteration
count $T$. For  $\varepsilon_{\text{total}}$ exceeding this scale,
further iterations no longer tighten the exploitability bound of
Corollary~\ref{thm:qcfr_bound} beyond the corresponding noise floor.
\end{proposition}

\subsection{Error-mitigation strategies}

Standard error-mitigation techniques apply directly to the QAE subroutine used inside Q-CFR: zero-noise extrapolation (running the amplitude-estimation circuit at several artificially amplified noise levels and extrapolating to the zero-noise limit), symmetry verification (using known game symmetries, e.g.\ card-relabeling invariance, to detect estimation outliers), postselection on auxiliary qubit measurement outcomes that indicate a failed amplitude-amplification step, and composite/dynamically-decoupled pulse sequences to suppress gate and crosstalk error during the Grover-diffuser layers. None of these techniques changes the asymptotic scalings above; they lower the constants inside $\varepsilon_{\text{gate}}$ and $\varepsilon_{\text{crosstalk}}$ and thereby push the crossover depth outward.

\section{Quantum Circuit Implementation}
\label{sec:circuits}

The QAE calls in Algorithm~\ref{alg:qcfr} are obtained via (i) a unitary state preparation
\begin{align}
   \ket{\psi}\ket{0}:= \hat{U}_{(I,a)}^{\sigma}\ket{0\ldots }\ket{0} = \sum_{s} \sqrt{p_{(I,a)}^{\sigma}(s)}\ket{s}\ket{0},
\end{align}
where $s$ denotes a trajectory, $p^{\sigma}_{(I,a)}(s)$ its probability in the normalized strategy-weighted counterfactual distribution, and (ii) an oracle $\hat{O}_f$, which encodes the (normalized) payoff of a trajectory as a rotation angle acting on the auxiliary qubit and controlled by the game-state register
\begin{align}
 \mathcal{\hat{O}}_f\ket{s}\ket{0}=\ket{s}\left(\sqrt{1-f(s)}\ket{0}+\sqrt{ f(s)}\ket{1}\right).
\end{align}
The composite system-auxiliary unitary $\mathcal{\hat{O}}_f(\hat{U}_{(I,a)}^{\sigma}
\otimes I)$ is then used to obtain the Grover operator, which is constructed following \cite{Brassard} as
$\hat{\mathcal{G}}_{(I,a)}^{\sigma}=\hat{S}_{\psi} \hat{S}_{\chi}$ where $\hat{S}_{\psi}$ is the reflection about the state $\mathcal{\hat{O}}_f\ket{\psi}\ket{0}$ and $\hat{S}_{\chi}$ is the reflection  
about the good subspace, where the auxiliary qubit is in state $\ket{1}$.

Whereas $\hat{S}_{\chi}$ is just a $Z$ gate acting on the auxiliary qubit, the flip $\hat{S}_{\psi}$ is conjugated to the flip about $\ket{0\dots0}$, which is realized as a multi-controlled $Z$ gate conjugated by $X$ gates. This can be decomposed further into elementary gates, for details see \cite{Barenco}.

QAE then estimates the probability of finding the auxiliary qubit in state $\ket{1}$, i.e.
\begin{align}
    \sum_s p_{(I,a)}^{\sigma}(s)f(s)
\end{align}
which is affinely related to the counterfactual value $ v^\sigma(I,a)$.

\section{Discussion}
\label{sec:discussion}

Q-CFR embeds quantum amplitude estimation into the regret-update loop of CFR, and Corollary~\ref{thm:qcfr_bound} shows that doing so preserves CFR's classical $\mathcal{O}(T^{-1/2})$ convergence rate up to an additive $\mathcal{O}(\varepsilon)$ term controlled entirely by the precision of the underlying QAE calls. Corollary~\ref{cor:speedup} makes explicit that this is a quadratic improvement in the \emph{per-counterfactual-value} sample complexity relative to classical Monte Carlo estimation, independent of any particular variance-reduction scheme.

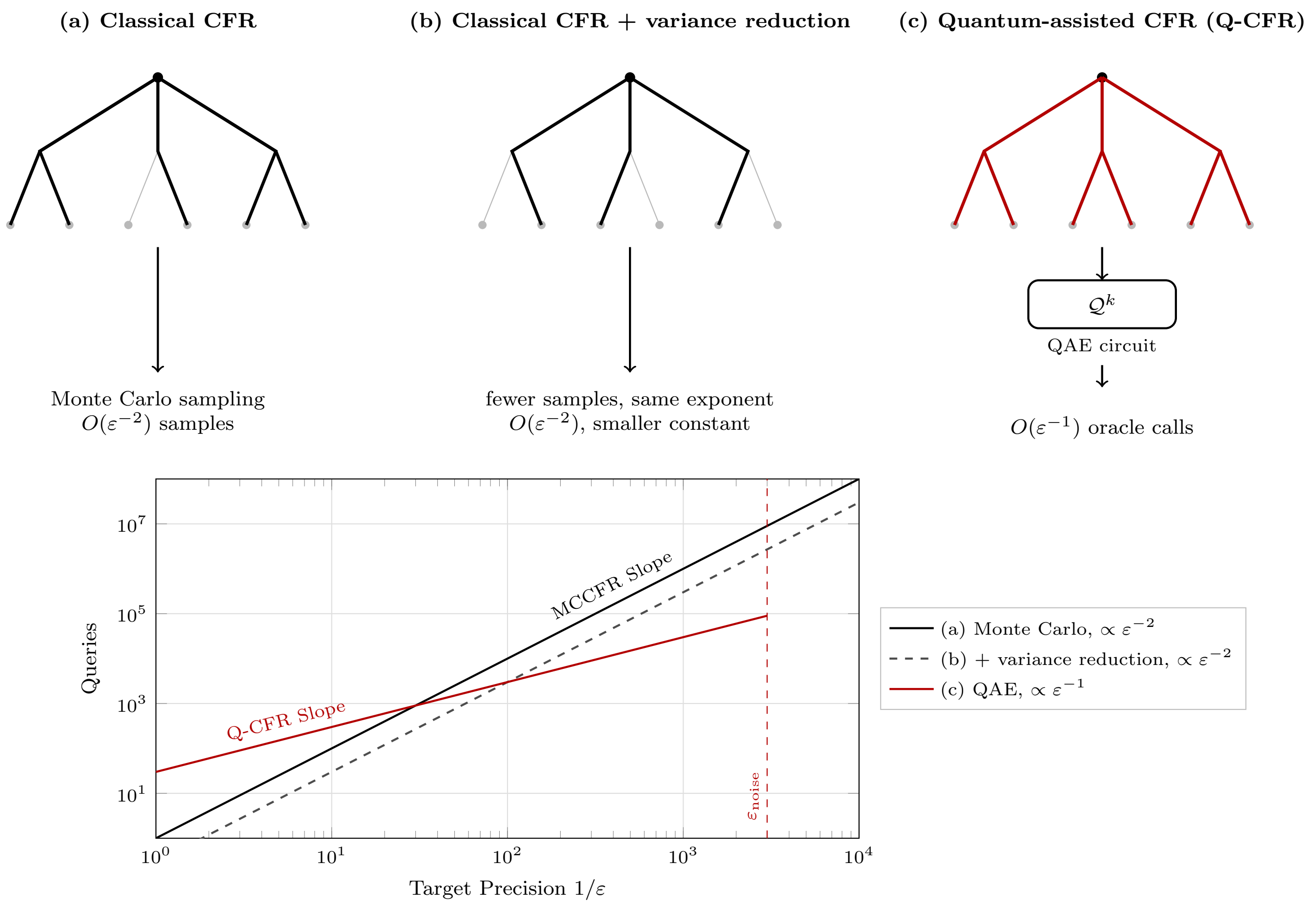
\begin{figure*}[h]
\centering
\begin{tikzpicture}[font=\footnotesize, x=1cm, y=1cm,
  edge/.style={gray!55, thin},
  samp/.style={very thick, black},
  qpath/.style={very thick, red!70!black}]

\newcommand{\gametree}[1]{%
  \draw[edge] (0,0) -- (-1.6,-1) (0,0) -- (0,-1) (0,0) -- (1.6,-1);
  \draw[edge] (-1.6,-1) -- (-2.0,-2) (-1.6,-1) -- (-1.2,-2)
              (0,-1) -- (-0.4,-2) (0,-1) -- (0.4,-2)
              (1.6,-1) -- (1.2,-2) (1.6,-1) -- (2.0,-2);
  \foreach \x in {-2.0,-1.2,-0.4,0.4,1.2,2.0} \fill[gray!55] (\x,-2) circle (1.6pt);
  \fill (0,0) circle (2pt);
}

\begin{scope}[shift={(0,0)}]
  \node at (0,0.75) {\textbf{(a) Classical CFR}};
  \gametree{}
  \draw[samp] (0,0) -- (-1.6,-1) -- (-2.0,-2);
  \draw[samp] (0,0) -- (-1.6,-1) -- (-1.2,-2);
  \draw[samp] (0,0) -- (0,-1) -- (0.4,-2);
  \draw[samp] (0,0) -- (1.6,-1) -- (1.2,-2);
  \draw[samp] (0,0) -- (1.6,-1) -- (2.0,-2);
  \draw[->, thick] (0,-2.3) -- (0,-4.0);
  \node[align=center] at (0,-4.55) {Monte Carlo sampling\\$O(\varepsilon^{-2})$ samples};
\end{scope}

\begin{scope}[shift={(6.4,0)}]
  \node at (0,0.75) {\textbf{(b) Classical CFR + variance reduction}};
  \gametree{}
  \draw[samp] (0,0) -- (-1.6,-1) -- (-1.2,-2);
  \draw[samp] (0,0) -- (0,-1) -- (-0.4,-2);
  \draw[samp] (0,0) -- (1.6,-1) -- (1.2,-2);
  \draw[->, thick] (0,-2.3) -- (0,-4.0);
  \node[align=center] at (0,-4.55) {fewer samples, same exponent\\$O(\varepsilon^{-2})$, smaller constant};
\end{scope}

\begin{scope}[shift={(12.8,0)}]
  \node at (0,0.75) {\textbf{(c) Quantum-assisted CFR (Q-CFR)}};
  \gametree{}

  \draw[qpath] (0,0) -- (-1.6,-1) -- (-2.0,-2);
  \draw[qpath] (-1.6,-1) -- (-1.2,-2);
  \draw[qpath] (0,0) -- (0,-1) -- (-0.4,-2);
  \draw[qpath] (0,-1) -- (0.4,-2);
  \draw[qpath] (0,0) -- (1.6,-1) -- (1.2,-2);
  \draw[qpath] (1.6,-1) -- (2.0,-2);
  \draw[->, thick] (0,-2.3) -- (0,-2.75);
  \draw[rounded corners, thick] (-1.0,-2.75) rectangle (1.0,-3.4);
  \node at (0,-3.075) {$\mathcal{Q}^{k}$};
  \node[font=\scriptsize] at (0,-3.65) {QAE circuit};
  \draw[->, thick] (0,-3.9) -- (0,-4.2);
  \node[align=center] at (0,-4.75) {$O(\varepsilon^{-1})$ oracle calls};
\end{scope}
\end{tikzpicture}

\vspace{1.2em}

\begin{tikzpicture}
\begin{axis}[
  width=0.62\textwidth, height=0.36\textwidth,
  xmode=log, ymode=log,
  xmin=1, xmax=1e4, ymin=1, ymax=1e8,
  xlabel={Target Precision $1/\varepsilon$},
  ylabel={Queries},
  font=\footnotesize, grid=major, grid style={gray!25},
  legend style={font=\scriptsize, at={(1.03,0.5)}, anchor=west, draw=gray!50},
  legend cell align=left, tick label style={font=\scriptsize}]
\addplot[thick, black, domain=1:1e4, samples=2] {x^2}
  node[pos=0.66, above, sloped, font=\scriptsize] {MCCFR Slope};
\addlegendentry{(a) Monte Carlo, $\propto\varepsilon^{-2}$}
\addplot[thick, dashed, gray!60!black, domain=1:1e4, samples=2] {0.3*x^2};
\addlegendentry{(b) + variance reduction, $\propto\varepsilon^{-2}$}
\addplot[thick, red!70!black, domain=1:3000, samples=2] {30*x}
  node[pos=0.22, above, sloped, font=\scriptsize] {Q-CFR Slope};
\addlegendentry{(c) QAE, $\propto\varepsilon^{-1}$}
\draw[dashed, red!70!black] (axis cs:3000,1) -- (axis cs:3000,1e8);
\node[red!70!black, font=\scriptsize, rotate=90, anchor=south west]
  at (axis cs:3000,1.6) {$\varepsilon_{\mathrm{noise}}$};
\end{axis}
\end{tikzpicture}
\caption{Schematic comparison of counterfactual-value estimation: (a) Monte Carlo estimation samples individual game-tree paths, (b) variance reduction needs fewer samples for the same precision but keeps the exponent, and (c) QAE uses all histories coherently inside one circuit and needs $O(\varepsilon^{-1})$ oracle calls. Bottom: samples or oracle calls versus target precision $1/\varepsilon$ on a log-log scale. Arbitrary prefactors are chosen so that the crossover is visible. The QAE curve ends at a noise floor $\varepsilon_{\mathrm{noise}}$ beyond which additional Grover iterations no longer improve the effective precision (Sec.~\ref{sec:noise}). The figure is a conceptual sketch of the scaling in Corollary~\ref{cor:speedup}, not measured data, and does not represent end-to-end CFR runtime.}
\label{fig:comparison}
\end{figure*}

This complements the classical variance-reduction techniques already used in large-scale CFR implementations, since they all reduce the constant in front of the classical $\mathcal{O}(\varepsilon^{-2})$ scaling, but cannot change the exponent of $\varepsilon$ itself as QAE does. The two approaches are in principle compositional: a classical variance-reduction baseline can be applied to the payoff function before it is loaded into the QAE oracle, further tightening the constant in the resulting $\mathcal{O}(\varepsilon^{-1})$ query count. Figure~\ref{fig:comparison} sketches this relationship schematically.

Two major limitations temper this picture.
First, the advantage established in Corollary~\ref{cor:speedup} is asymptotic and assumes idealized (noiseless) QAE calls.

Second, and most importantly for the present manuscript, the noise considerations and the qualitative crossover analysis of Sec.~\ref{sec:noise} are analytic projections from the underlying QAE literature rather than measurements on real quantum hardware or even a full noisy-simulator study of Q-CFR itself.

\section{Conclusion \& Outlook}
\label{sec:conclusion}

With our findings, we state that: if a payoff oracle of the assumed form is available at acceptable cost and the amplitude-estimation calls run at sufficient depth, then the estimation component of equilibrium computation can be accelerated without sacrificing CFR's guarantees. Whether this translates into an end-to-end advantage over sampling-based CFR variants for games of practical size depends on the oracle cost, the noise floor of the hardware, and the variance properties of the classical baseline.

We consider formulating these open questions as concrete experiments: Running Algorithm~\ref{alg:qcfr} against a noisy simulator and, subsequently, real NISQ hardware on a benchmark game such as Kuhn or Leduc poker, and comparing measured exploitability curves against both classical CFR and the bound of Corollary~\ref{thm:qcfr_bound} as the essential next step, and we deliberately do not report placeholder numerical results instead.
Their outcome will decide whether Q-CFR remains a theoretical template or becomes a practical tool for the auction, security, and negotiation settings that motivate this work.
Beyond these technical steps, the broader motivation for this line of work is that many practically important decision problems, such as automated negotiation, dynamic auctions, and security-resource-allocation games, are naturally graded-reward, imperfect-information games for which even modest reductions in the sample complexity of equilibrium computation could be practically significant.

\section*{Acknowledgements}
We thank U.~Armbrüster, M.~Bauer, S.~Edelkamp, S.~R.~Maschek, J.~Schwitalla, D.~Smith, and J.~Wehling for helpful discussions.

\section*{Declaration of AI Assistance} The authors used Claude for text drafting and stylistic revision in the introductory sections, and for sketching the schematic comparison. All technical statements, citations, and mathematical formulations were independently checked and approved by the authors.
\bibliographystyle{unsrt}
\bibliography{references_arxiv}

\appendix

\section{Circuit-Level Implementation Details}
\label{app:circuits}

\subsection{Resource considerations}

For a fixed strategy $\sigma^t$, the QAE subroutine of Algorithm~\ref{alg:qcfr} estimates the expectation of a bounded, normalized payoff. The quantum resources are determined by three components:

\begin{enumerate}[label=(\roman*)]
    \item{A state-preparation circuit $\hat{U}$ that encodes the distribution in a register.}
    \item{A payoff oracle $\hat{O}_f$ that loads $f(s)$ as a rotation angle on a single qubit.}
    \item{A Grover operator $\hat{\mathcal{G}}$ built from $\hat{U}$ and $\hat{O}_f$, which is applied up to $\mathcal{O}(\varepsilon^{-1})$ times per estimate.}
\end{enumerate}

Because $\sigma^t$ changes in every iteration, $\hat{U}$ must be re-parameterized in every iteration. The cost of $\hat{U}$ and $\hat{O}_f$ depends
on the game and its encoding and is multiplied by the number of Grover
applications. For large games we therefore expect it to dominate the overall
cost. IQAE and ML-AE avoid the $\mathcal{O}(\log(1/\varepsilon))$ phase-estimation
auxiliary qubits of canonical QAE \cite{Grinko2019,Suzuki2019}, but the size of the
history register and the cost of building $\hat{U}$ and $\hat{O}_f$ are instance-specific.
A reliable estimate requires an explicit circuit construction and compilation for a concrete game, which we leave to future work (Sec.~\ref{sec:conclusion}).

\subsection{Transpilation considerations}

For gate-model hardware, mapping the logical Q-CFR circuits above to a physical device involves

\begin{enumerate}[label=(\roman*)]
    \item{Logical-level optimization: commuting-gate cancellation, single-qubit gate merging, and algebraic simplification of the rotation angles used to encode counterfactual values.}
    \item{Physical-level mapping via a qubit-routing heuristic that minimizes \textit{SWAP} overhead for the multi-controlled operations appearing in the Grover operator, expressed in the device's native gate set. }
\end{enumerate}

\section{NISQ-Friendly QAE Variants: Additional Detail}
\label{app:nisq}

This appendix collects implementation-level detail on the two QAE variants used in Sec.~\ref{sec:nisq}; the complexity claims themselves, and their proofs, are established in the cited original works and are only summarized here for self-containedness.

\textbf{IQAE} \cite{Grinko2019} maintains a running confidence interval $[\theta_{\min}^{(i)}, \theta_{\max}^{(i)}]$ on the amplitude-encoding angle $\theta$ after each round $i$, and chooses the next Grover power $k_{i+1}$ to be the largest power such that the images of $\theta_{\min}^{(i)}$ and $\theta_{\max}^{(i)}$ under the $k_{i+1}$-fold-amplified circuit remain distinguishable given the per-round shot budget. This adaptive doubling-type schedule yields the $\mathcal{O}(\varepsilon^{-1})$ depth scaling without requiring the $\mathcal{O}(\log(1/\varepsilon))$ phase-estimation auxiliary qubits of canonical QAE.

\textbf{ML-AE} \cite{Suzuki2019} instead fixes the schedule of Grover powers in advance (typically geometric, $k_j = 2^{j}$ for $j=0,\dots,m$), collects measurement counts at each depth, and recovers $\theta$ (equivalently the amplitude) via classical maximum-likelihood estimation over the resulting binomial likelihoods; the fixed schedule trades circuit-scheduling flexibility for predictable, hardware-plannable resource usage across many parallel counterfactual-value queries.

\end{document}